\documentclass[twocolumn,a4paper]{article}
\usepackage[textwidth=0.87\paperwidth,textheight=0.8\paperheight]{geometry}
\usepackage[affil-it]{authblk}

\usepackage[tt=false, type1=true]{libertine}
\usepackage{booktabs}

\usepackage{hyperref}
\usepackage{graphicx}
\graphicspath{{../pdf/}{../jpeg/}}
\DeclareGraphicsExtensions{.pdf,.jpeg,.png}

\usepackage[cmex10]{amsmath}
\usepackage{amssymb}

\usepackage{url}
\usepackage{xspace}
\usepackage{todonotes}

\usepackage[switch,columnwise]{lineno}

\usepackage[ruled,linesnumbered]{algorithm2e}
\usepackage{xcolor}
\usepackage{tikz}
\usepackage{pgfplots}
\pgfplotsset{compat=1.13}
\usepackage{numprint}
\npdecimalsign{.}

\usepackage{subcaption}
\usepackage{placeins}
\usepackage[normalem]{ulem}

\newcommand{\tool}[1]{{\small\textsf{#1}}}
\newcommand{\parmetis}{\tool{ParMetis}\xspace}

\newcommand{\kahip}{\tool{KaHIP}\xspace}
\newcommand{\bakpa}{\tool{Geographer}\xspace}
\newcommand{\zoltan}{\tool{Zoltan}\xspace}
\newcommand{\rib}{\tool{RIB}\xspace}
\newcommand{\rcb}{\tool{RCB}\xspace}
\newcommand{\mj}{\tool{MultiJagged}\xspace}
\newcommand{\zsfc}{\tool{zoltanSFC}\xspace}

\newcommand{\dist}{\operatorname{dist}}
\newcommand{\ie}{i.\,e.,\xspace}
\newcommand{\eg}{e.\,g.,\xspace}
\newcommand{\etal}{et al.\xspace}

\newcommand{\wrt}{w.\,r.\,t.\xspace}
\newcommand{\NN}{\mathbb{N}}
\newcommand{\RR}{\mathbb{R}}
\newcommand{\graphname}[1]{{\texttt{#1}}}

\definecolor{markedcolor}{RGB}{31,120,180}
\definecolor{plottinggreen}{RGB}{178,223,138}
\definecolor{thirdhue}{RGB}{228,26,28}
\definecolor{parMetis}{RGB}{190,174,212}
\definecolor{parHip}{RGB}{120,16,100}

\colorlet{zoltanMJ}{thirdhue}
\colorlet{zoltanRcb}{orange}
\definecolor{zoltanRib}{RGB}{190,174,212}
\colorlet{zoltanHsfc}{plottinggreen}
\definecolor{geoKmeans}{RGB}{31,120,180}

\newcommand{\hide}[1]{\textcolor[rgb]{0.2,0.2,0.2}{[#1]}\xspace}
\renewcommand{\hide}[1]{}

\title{Balanced k-means for Parallel Geometric Partitioning}

\author{Moritz von Looz}
\author{Charilaos Tzovas}
\author{Henning Meyerhenke}
\affil{University of Cologne\\ \{mloozcor, ctzovas, h.meyerhenke\}@uni-koeln.de}

\begin{document}


\maketitle

\begin{abstract}
Mesh partitioning is an indispensable tool for efficient parallel numerical simulations. 
Its goal is to minimize communication between the processes of a simulation while achieving load balance.
Established graph-based partitioning tools yield a high solution quality; however, their scalability is limited.
Geometric approaches usually scale better, but their solution quality may be unsatisfactory for ``non-trivial'' mesh topologies.

In this paper, we present a scalable version of $k$-means that is adapted to yield balanced clusters.
Balanced $k$-means constitutes the core of our new partitioning algorithm \bakpa.
Bootstrapping of initial centers is performed with space-filling curves, leading to fast convergence of the subsequent balanced k-means algorithm.

Our experiments with up to \numprint{16384} MPI processes on numerous benchmark meshes show 
the following: (i) \bakpa produces partitions with a lower communication volume than state-of-the-art geometric partitioners from the \zoltan package; 
(ii) \bakpa scales well on large inputs;
(iii) a Delaunay mesh with a few billion vertices and edges can be partitioned in a few seconds.
\end{abstract}

\section{Introduction}
\label{sec:introduction}
%
In simulations of spatial phenomena, it is common to discretize the simulation domain into a geometric graph called \emph{mesh}.
In the common use case of modeling with partial differential equations (PDEs)~\cite{MortonM05numerical}, this discretization ultimately leads to linear systems or explicit time-stepping methods.
The resulting matrices are typically very large and sparse, requiring parallelization for efficient solutions.
To optimize sparse matrix-vector multiplication (SpMV, often the major computational kernel in these simulations), in particular its communication, one needs to distribute
the mesh onto the processing elements (PEs) such that (i) the load is balanced and (ii)
the communication between PEs is minimized.

A common strategy for computing such a distribution is to solve the graph partitioning
problem~\cite{SchloegelKarypisKumar02graph,bichot2013graph,SPPGPOverviewPaper} for the primal
or dual graph of the mesh. 
Its most common formulation for an undirected graph $G=(V,E)$ asks for
a division of $V$ into $k$ pairwise disjoint subsets (\emph{blocks}) such
that all blocks are no larger than $(1+\epsilon) \cdot \left\lceil \frac{|V|}{k} \right\rceil$ (for small $\epsilon \geq 0$) 
and some objective function modeling the communication volume is minimized. Traditionally, the \emph{edge cut}, \ie the total number of edges having their incident vertices in different blocks, is used as a proxy for the communication volume -- despite some drawbacks~\cite{HendricksonK00graph}.

\subsubsection*{Motivation}
Established tools for general-purpose graph partitioning typically yield a high quality in terms of the edge cut.
This is mainly due to the effectiveness of the multilevel approach~\cite{SPPGPOverviewPaper}.
On the other hand, since this approach uses a hierarchy of successively smaller graphs, its scalability
shows some limitations. Previous work often saw an increase in running time when moving beyond a few
hundred PEs for such approaches~\cite{HoltgreweSS10engineering,DBLP:conf/sc/KirmaniR13,DBLP:journals/tpds/MeyerhenkeSS17}.
Thus, for large-scale simulations the research community has moved to more scalable geometric
methods, \eg space-filling curves~\cite{Bader:2012:SCI:2412036}, or seemingly simpler space-partitioning methods~\cite{DRDC16}. 
While geometric methods are often much faster and more scalable than graph-based ones, their partitioning quality
is typically worse, leading to a higher running time of the targeted application.

Good block shapes (connected, compact, to some extent convex) are not only beneficial for certain applications~\cite{DiekmannPreisSchlimbachWalshaw00shape}; 
often they also come along with a high partitioning quality \wrt established graph metrics~\cite{MeyerhenkeMS09new}.
Graph-based tools are usually not satisfactory in this regard unless specifically designed for this purpose~\cite{MeyerhenkeSchamberger06parallel}.

While most simulations are nowadays performed in 3D, the same does not hold for all mesh partitioning problems.
%
%
Indeed, two key application areas for massively parallel PDE solvers are the atmosphere and ocean simulations in weather and climate models; they feature prominently on the list of exascale challenges~\cite{Sarkar2009}.
Although the simulations are run in 3D, their vertical extent is typically very small and variable over the application domain.
Thus, the mesh tends to be partitioned in 2D and then extended to a 3D mesh during the simulation using topography information;
therefore this type of mesh/problem is sometimes called 2.5-dimensional.
The computational effort depends on the number of 3D grid points and is reflected in the 2D mesh as a node weight.

Following from these requirements, we are interested in a scalable mesh partitioning algorithm for 2D and 3D meshes that yields high quality in terms of block shapes \emph{and} relevant graph metrics.

\subsubsection*{Contribution}
We present a new parallel algorithm for direct $k$-way partitioning of geometric point sets corresponding to simulation meshes.
Our algorithm and its implementation are called \bakpa for \emph{\underline{Geo}metry-based \underline{graph} partition\underline{er}} (Section~\ref{sec:algorithm:sub:kmeans}).
It consists of two phases: the first one initially partitions the input points based on a space-filling curve,
the second one is based on Lloyd's popular $k$-means algorithm, which is known to produce convex block shapes. 
We add, however, a weighting scheme to obtain balanced block sizes.
With some geometric optimizations adapted to the weighting scheme, including distance bounds of Hamerly \etal~\cite{hamerly2010making}, our algorithm scales to large inputs and a reasonably large number of MPI processes.
For example, using \numprint{16384} processes, a mesh with 2 billion vertices can be partitioned into \numprint{16384} blocks within a few seconds.
The partitions derived this way have good global shapes and a low communication volume.

Our experiments (Section~\ref{sec:experiments}) on a variety of meshes indicate a quality at least competitive with state-of-the-art parallel geometric partitioning methods in \zoltan and \parmetis.
The total communication volume of the generated partitions is on average 15\% lower than that of the best competitor (MJ); this advantage is most pronounced for 2D meshes.
While not all established graph partitioning metrics can be improved by \bakpa for all instance classes,
the average SpMV communication time is reduced in all classes compared to \zoltan's geometric partitioners.
Moreover, our scaling behavior is similar to the best competitor and better than recursive methods.

\section{Problem Definition}
\label{sec:prob-def}
%
Depending on how the mesh defines data and its dependencies, load balancing
by partitioning may work with the coordinate set, the mesh itself or its dual graph.
In general, a \emph{graph partitioning} technique focuses on the graph information may or may not use the geometric information, \ie the coordinates of the vertices of the graph to be
partitioned. A \emph{geometric partitioning} technique focuses on the coordinate information.
In this paper, we present a new geometric partitioning algorithm and compare it against previous work in that area - but to evaluate its impact on mesh partitioning applications, we measure the quality of results also in graph-based metrics.
Note that a graph-based postprocessing, for example based on the Fiduccia-Mattheyses local refinement heuristic is easily possible, but outside the scope of this paper.

In its general-purpose form without geometric information, the graph
partitioning problem (GPP) is defined as follows: Given a number $k \in \NN_{>1}$ and an undirected graph $G=(V,E,\omega)$ with 
$n := \vert V \vert$, $m := \vert E \vert$, and non-negative edge weights $\omega: E \to \RR_{>0}$, 
find a partition $\Pi$ of $V$ with \emph{blocks} of vertices $\Pi = (V_1$, \dots, $V_k)$ such that some objective function is optimized.
A \emph{balance constraint} requires that all blocks must have approximately equal size (or weight). More precisely, it requires 
that, $\forall i\in \{1, \dots, k\}: |V_i|\leq L_{\max} := (1+\epsilon)\lceil |V|/k \rceil$ for
some imbalance parameter $\epsilon \in \RR_{\geq 0}$.
A block $V_i$ is \emph{overloaded} if $|V_i| > L_{\max}$. Note that the balance constraint definition can be extended to vertex-weighted graphs~\cite{HendricksonL95improved}.

How to measure the quality of a partition depends mostly on the application. 
A common metric is the \emph{edge cut}, \ie the total number (or summed weights) of all edges whose incident vertices are in different blocks.
Besides the edge cut, we also include the maximum and total \emph{communication volume}, 
as they measures communication costs in parallel numerical simulations more accurately~\cite{HendricksonK00graph} 
To assess the partition shapes, we include results on the partition diameter, in many spatial simulations the diameter of a block influences the efficiency of the computation.
The measures for a block $V_i \in \Pi$ are defined as:

\begin{align*}
&\text{external edges or cut-edges:}\\
ext(V_i) :&= |\{e=\{u,v\}\in E:u\in V_i\wedge v\not\in V_i\}|,\\
&\text{communication volume:}\\
comm(V_i) :&= \sum_{v\in V_i} |\{V_j\in\Pi : \exists u \in V_j \wedge \{u,v\}\in E\}| \\
&\text{diameter:}\\
diam(V_i) :&= \max_{u,v\in V_i} \dist(u,v).
\end{align*}

Note that in unweighted graphs, the edge cut is the summation norm of the external edges
divided by $2$ to account for counting each edge twice.

To measure the quality of a partition empirically,
we redistribute the input graph according to it, perform sparse matrix-vector multiplications (SpMVs) with the adjacency matrix and a suitable chosen vector
and measure the communication time needed within the SpMV.
To average out random fluctuations, we average the time over 100 multiplications on each instance.

\section{Related Work}
\label{sec:rel-work}
%
%
A comprehensive review of the state of the art is beyond the scope of this paper. The interested reader is referred to survey articles~\cite{SchloegelKarypisKumar02graph,SPPGPOverviewPaper} and a book~\cite{bichot2013graph}. 
We focus in our description mostly on closely related (parallel, geometric, shape-optimizing) established techniques and tools, in particular those used in the experimental evaluation.

A related approach in mesh partitioning is to consider only the graph structure and not (necessarily) its geometry.
Probably the most popular among these general graph partitioners is the parallel tool \parmetis~\cite{SchloegelKK02parallel},
which is based on the Fiduccia-Mattheyses (FM) heuristic and particularly appreciated for its fast running time.

Other parallel graph partitioners include \tool{PT-Scotch}~\cite{Pellegrini12scotch} and the parallel versions of \tool{Jostle}~\cite{Walshaw-MPDD-07}, \tool{DibaP}~\cite{Meyerhenke12shape},  and \kahip~\cite{DBLP:journals/tpds/MeyerhenkeSS17}.
These tools follow the multilevel approach; they construct a hierarchy of successively smaller graphs.
While yielding high solution quality, this approach seems to be the main bottleneck for high scalability, though.
Avoiding this drawback, the tool \tool{xtraPulp}~\cite{xtrapulp} uses distributed label propagation.
It is, however, targeted at complex networks and pays its improved scalability with a quality penalty.
%
    
\subsection{Geometric techniques}
Most geometric partitioning techniques in wide use consider points and optimize for load balance ~\cite{DRDC16, MTHAS16,SIMON1991135}.
Established geometric methods include the recursive coordinate bisection (\rcb~\cite{SIMON1991135,BergBokh87}) and recursive inertial bisection (\rib~\cite{Taylor_Nour,Williams_91}).
Deveci \etal~\cite{DRDC16} introduce a multisection algorithm, called Multi\-Jagged (MJ), as a generalization of the traditional recursive techniques.
The space is divided into rectangles while minimizing the weight of the largest rectangle.
This method has better running time and is more scalable  but yields less balanced partitions compared to RCB.
Many common geometric partitioning methods are implemented in the \tool{Zoltan} toolbox~\cite{zoltan}.

Another class of geometric techniques uses space-filling curves (SFCs), usually the Hilbert curve~\cite{Bader:2012:SCI:2412036}.
These techniques are also fast and scalable and rely on the fact that two points whose indices on the curve are close, are also often close in the original space.
While load balance is fairly easy to maintain, the quality of the computed partitions in terms of graph-based methods is relatively poor for non-trivial meshes~\cite{HW02}.
Implementations of partitioning algorithms using space-filling curves are available in the \parmetis and \zoltan packages.


%


\subsection{Shape optimization}
The benefit of optimizing block shapes has been acknowledged in a number of publications, not only for certain applications~\cite{DiekmannPreisSchlimbachWalshaw00shape}, but also for established graph metrics~\cite{MeyerhenkeMS09graph}. 
However, previous shape-optimizing approaches suffer from a relatively high running time for static partitioning~\cite{MeyerhenkeMS09new,MeyerhenkeMS09graph,DBLP:journals/jcphy/FuLHA17} and limited
scalability~\cite{DiekmannPreisSchlimbachWalshaw00shape,Meyerhenke12shape}. 


The \tool{bubble framework} introduced by Diekmann \etal~\cite{DiekmannPreisSchlimbachWalshaw00shape} achieves well-shaped partitions by repeatedly selecting center vertices and growing blocks around them using constrained breadth-first search.
This concept is similar to the $k$-means problem (cf.\ Section~\ref{subsub:rel-work-k-means} below), except that cluster mem\-ber\-ship and centers are computed with graph-theoretic instead of geometric distances.
Due to the discrete nature of graph distances, the center selection can be computationally demanding.

Shape optimization with \tool{Bubble-FOS/C}~\cite{MeyerhenkeMS09graph}, a variation of the \tool{bubble framework} with diffusion distances for the part resembling $k$-means, has been shown to have some theoretical foundation~\cite{DBLP:journals/algorithmica/MeyerhenkeS12}: 
similar to spectral partitioning, it computes the global optimum of a relaxed edge cut optimization problem.
At the same time, the quality of diffusive partitioning is in practice typically higher than that of spectral methods.

\subsection{$k$-Means}
\label{subsub:rel-work-k-means}
The $k$-means problem, common in unsupervised machine learning and clustering, consists of a set of points $P$ in a metric space $\mathbb{H}$ and a number $k$ of target clusters.
It asks for an assignment of the points in $P$ to $k$ clusters so that the sum of squared distances of each point to the mean of its cluster is minimized.
This target function bears no relation to our graph-based metrics, but local minima yield Voronoi diagrams, whose sides are convex,
useful for our geometric partitioning phase with shape optimization.

Lloyd's greedy algorithm~\cite{Lloyd82least} for the $k$-means problem consists of repeatedly alternating two steps:
\begin{itemize}
\item For every point $p\in P$: Assign $p$ to the cluster $c$ so that the distance between $p$ and $\mathrm{center}(c)$ is minimized.
\item For every cluster $c$: Set cluster center $\mathrm{center}(c)$ to the arithmetic mean of all points in $c$.
\end{itemize}
The algorithm stops when the maximum movement of cluster centers is below a user-defined threshold, at the latest if no cluster membership changes.

In each step, the sum of squared distances between each point and the center of its cluster decreases. As distances are nonnegative, the algorithm eventually converges to a local optimum.

Which local optimum is reached, depends on the choice of initial centers.
A straightforward option is to choose them uniformly at random, with erratic and arbitrarily bad results~\cite{arthur2007k}.
Alternatives include K-Means++~\cite{arthur2007k}, which chooses the first center at random and then iteratively chooses each subsequent center to maximize the distance to all existing centers.
Unfortunately, this method is inherently sequential and the complexity of $\mathcal{O}(nk)$ required by $k$ passes over $n$ points is too expensive for our scenario.

Bachem et al~\cite{bachem2016fast} present a probabilistic seeding method using \emph{Markov Chain Monte Carlo} (MCMC) sampling and claim an effective complexity of $\mathcal{O}(n+k)$ for similar quality.
Still, the empirical running times are in the order of minutes for a few million points~\cite{bachem2016fast}.

Several approaches exist to accelerate the main phase of $k$-means, many of them exploiting the triangle inequality.
Elkan~\cite{elkan2003using} keeps one upper bound and $k$ lower bounds for each data point $p$, avoiding distance calculations to clusters that cannot possibly be the closest one.
Hamerly et al.~\cite{hamerly2010making} simplify this approach to one upper bound for the distance of each point to its own center and one lower bound for the distance to the \emph{second-closest} center. When for a point $p$ this first bound is below the second, the cluster membership of $p$ cannot have changed and the loop over all centers can be skipped. In both works, the bounds are relaxed when the cluster centers move and are updated to their exact values when a distance calculation becomes necessary.

Sculley~\cite{sculley2010web} presents a sampling method for $k$-means using gradient descent, with a claimed speedup of two orders of magnitude over methods using the triangle inequality. 
Unfortunately, his method does not fit well for parallelization or for extension with a balance constraint.

%


\section{Weighted Balanced $k$-Means for Mesh Partitioning}
\label{sec:algorithm:sub:kmeans}
The input for $k$-means commonly consists of a set of points $P$ and a target cluster count $k$.
We also accept a balance parameter $\epsilon$ and an optional weight function $w: P \rightarrow \mathbb{R}^+$.
In the unweighted case, we set each vertex weight to one.

The objective we use is then similar to the graph partitioning problem:
Find an assignment of points to blocks so that the weight sum of each block is at most $1+\epsilon$ times the average weight sum and the sum of squared point-center distances is minimized.\footnote{When non-uniform block sizes are desired, for example when partitioning for heterogeneous architectures, this can easily be adapted. However, it is not the focus of this work.}
This is $\mathcal{NP}$-hard, as it contains the classical $k$-means problem.

Starting from Lloyd's algorithm, we discuss the changes we made to address parallelization, balancing and geometric optimizations, finally presenting the overall algorithm.

\subsection{Parallelization and Space-Filling Curves}
\label{subsubsec:k-means-parallel}
Lloyd's algorithm parallelizes well and our extensions do not change that.
Each processor stores a subset of the points, while the cluster centers and influence values are replicated globally.
The computationally most expensive phase is assigning points to the appropriate cluster, which can be done independently for each point.
After points are assigned, a parallel sum operation is performed to calculate the new cluster centers and sizes.

As preparation for the geometric optimizations, we globally sort and redistribute all points according to their index on a space-filling curve,
thus ensuring that each processor has local points that are grouped spatially and their bounding box is reasonably tight.
For this distributed sorting step, we use the scalable quicksort implementation of Axtmann \etal~\cite{axtmann2017quicksort}.

\subsection{Balancing}
\label{subsubsec:balancing}
To achieve balanced cluster sizes, we add an \emph{influence} value to each cluster, initialized to 1.
In the assignment phase, instead of assigning each point $p$ to the cluster with the smallest distance,
we assign it to the cluster $c$ for which the term $\dist (p, \mathrm{center}(c)) / \mathrm{influence}(c) $ is minimized.
We call this term the \emph{effective distance} of $p$ to $\mathrm{center}(c)$.
This approach results in the creation of \emph{weighted Voronoi diagrams}~\cite{aurenhammer1984optimal} (which are not necessarily convex).

After all points are assigned, the global weight sum is calculated for each block.
The influence value of oversized blocks is decreased, of undersized blocks increased.
How strongly to increase or decrease the influence in response to an imbalanced partition is a tuning parameter.
Our decision is guided by geometric considerations:
The volume of a $d$-dimensional hypersphere with radius $r$ scales with $r^d$.
Assuming a roughly uniform point density, increasing the effective distance of a cluster to all points by a factor of $b$ leads, all else being equal,
to a change in size of $b^{-d}$.
Thus, if the ratio of the target size and current size for a cluster $c$ is $\gamma(c)$, we set:
\begin{equation}
\mathrm{influence}[c] \gets \mathrm{influence}[c]/\gamma(c)^{1/d}. 
\label{eq:influence}
\end{equation}

Then, the new expected size of cluster $c$ is $\left(\frac{1}{\gamma(c)^{1/d}}\right)^{-d} \cdot \mathrm{size}_\mathrm{old}$ =
$\left(\gamma(c)^{1/d}\right)^d \cdot \mathrm{size}_\mathrm{old}$ = $\gamma(c) \cdot \mathrm{size}_\mathrm{old}$ = $\mathrm{size}_{\mathrm{target}}$.

Of course, the input points are usually not uniformly distributed and more than one balance iterations is needed.
To prevent oscillations, we restrict the maximum influence change in one step to $5\%$.
This approach is repeated for a maximum number of balancing steps or until the maximum imbalance is at most $\epsilon$; then centers are moved and a new assign-and-balance phase starts.
The maximum number of balancing iterations between center movements is a tuning parameter.

In very heterogeneous point distributions, it can happen that clusters need very small or very large influence values to gain a reasonable size.
If, after a movement phase, cluster centers with very different influence values are in close proximity, anomalies such as empty or absurdly large clusters might occur.
To avoid such cases, we add an \emph{influence erosion} scheme:
When cluster centers move, we regress their influence value according to a sigmoid function of the moved distance.
Let $\delta(c)$ be the distance that $\mathrm{center}(c)$ moved in the last phase and let $\beta(C)$ be the average cluster diameter.
We define an \emph{erosion factor} $\alpha(c)$ between 0 and 1, controlling how strongly the influence is eroded. 
Then:

\begin{align}
\alpha(c) = \frac{2}{1+\exp(\min (-\delta(c)/\mathit{\beta(C)}, 0)    )    } - 1\\
\mathrm{influence}(c) \gets \exp((1-\alpha(c))\cdot\log(\mathrm{influence}(c)))
\end{align}

After moving more than the average distance between centers, the influence value is thus almost back to 1, as an influence appropriate for one neighborhood might not be appropriate for a different neighborhood (of clusters).

\subsection{Geometric Optimizations}
\label{subsubsec:k-means-geometric}
We adapt the distance bounds of Hamerly et al.~\cite{hamerly2010making} for effective distances.
Let $p$ be a point and $c := c(p)$ its assigned cluster; then $\mathrm{ub}(p)$ stores an upper bound for the effective distance of $p$ and $c$; $\mathrm{lb}(p)$ stores a lower bound for the second-smallest effective distance.
If $\mathrm{lb}(p) > \mathrm{ub}(p)$ holds when evaluating the new cluster assignment of point $p$, it is still in its previous cluster and distance calculations to other clusters can be skipped.

When a cluster center moves or its influence value changes, these bounds need to be relaxed to stay valid.
Again, let $\delta(c)$ be the distance that $\mathrm{center}(c)$ moved in the last phase.
For each point $p$ in cluster $c$, the new upper bound $\mathrm{ub}'(p)$ is then:
\begin{equation}
\mathrm{ub}'(p) = \mathrm{ub}(p)-\delta(c)/\mathrm{influence}(c(p)).
\label{eq:ub-influence}
\end{equation}
The lower bounds are relaxed with the maximum combination of $\delta$ and influence, as any cluster could be the second-closest one.
\begin{equation}
\mathrm{lb}'(p) = \mathrm{lb}(p)+\max_{c'\in C} \delta(c')/\mathrm{influence}(c')
\label{eq:lb-influence}
\end{equation}
Using these bounds, the innermost loop can be skipped in about 80\% of the cases, more in the later phases where centers and influence values change less.
Nearest-neighbor data structures like kd-trees are outperformed by simpler distance bounds in most published experiments~\cite{drake2012accelerated,hamerly2010making}.

\subsection{Bounding Boxes}
%
For a given point $p$, most clusters are unlikely candidates. In fact, the likely cluster centers lie roughly in a $d$-dimensional hypersphere around $p$.
By calculating the bounding box around the process-local points and sorting the cluster centers by their effective distance to it,
we can stop evaluating clusters for a point $p$ when their minimum effective distance is above the ones for already found candidates.

\subsection{Algorithm}
\begin{algorithm}
\KwIn{centers $C$, local points $P_{\text{local}}$, node weights $W$, \ensuremath{\mathrm{influence}}, previous assignments $A_{\mathrm{old}}$, ub, lb, $\epsilon$}
\KwOut{assignments $A$, new \ensuremath{\mathrm{influence}}, bounds ub, lb}
  bb $\gets$ bounding box around local points $P_{\text{local}}$\;
 
 \For{center $c \in C$}{
  distToBb[$c$] $\gets$ maxDist(bb,$c$)$/\mathrm{influence}[c]$\;
  localBlockSizes[$c$] $\gets$ 0\;
 }
 sort centers $C$ by distToBb\;\label{line:sort-centers}
 \For{$i \in \{0, ..., \mathrm{maxBalanceIter}\}$}{
 \For{$p\in P_{\text{local}}$}{\label{line:local-loop}
    \If{ub[$p] < $lb[$p$]}{
      $A[p] \gets A_{\mathrm{old}}[p]$\;\label{line:hamerly-optimization}
    }
    \Else{
      bestValue, secondBestValue $\gets$ $\infty$\;
      \For{$c \in C$}{
      \If{$\text{distToBb}[c] > $secondBestValue}{
	break\;\label{line:bb-optimization}
      }
      effDist = dist($c,p) / \mathrm{influence}[c]$\;
      \uIf{effDist$ < $bestValue}{
       $A[p] \gets c$\;\label{line:assign-cluster}
       secondBestValue $\gets$ bestValue\;
       bestValue $\gets$ effDist\;
      }
      \uElseIf{effDist$ < $secondBestValue}{
	secondBestValue $\gets$ effDist\;
      }
      }
      ub[$p$] $\gets$ bestValue\;\label{line:tighten-ub}
      lb[$p$] $\gets$ secondBestValue\;\label{line:tighten-lb}
      }
      
      localBlockSizes$[A[p]] += W[p]$\;
    }    
    \textcolor{markedcolor}{globalSizes $\gets$ globalSumVector(localBlockSizes)}\;\label{line:communicate-sizes}
    \If{imbalance(globalSizes) $< \epsilon$}{
      \Return{$A$, $I$, ub, lb}
    }
    \For{$c \in C$}{
    $\mathrm{influence}(c) \gets$ adaptInfluence($\mathrm{influence}(c)$, globalSizes[$c$]); \tcc*[f]{Eq.~(\ref{eq:influence})}\label{line:adapt-influence}
    }
    update ub\;
    update lb\;
  }
 \Return{$A$, $I$, ub, lb}
 \caption{AssignAndBalance}
 \label{algo:assign-and-balance}
\end{algorithm}

Algorithm~\ref{algo:assign-and-balance} shows the resulting assign-and-balance phase of our $k$-means and is executed by all processors in parallel.
The first~\ref{line:sort-centers} lines prepare data structures and optimizations, the main loop starts in Line~\ref{line:local-loop}.
If the distance bounds for a point $p$ guarantee that its cluster assignment has not changed, the inner loop can be skipped (Line~\ref{line:hamerly-optimization}).
If not, we iterate over the cluster centers and assign $p$ to the one with the smallest effective distance (Line~\ref{line:assign-cluster}).
As soon as the effective distance between a cluster center and the bounding box of local points is higher than the second best value found so far,
the remaining clusters cannot improve on that and can be skipped (Line~\ref{line:bb-optimization}).

We update bounds for all points where distance calculations were necessary (Lines~\ref{line:tighten-ub} and \ref{line:tighten-lb}).
Finally, global block sizes are computed as sums of all local block sizes (Line~\ref{line:communicate-sizes}).
This is the only part requiring communication in the balance routine, marked in \textcolor{markedcolor}{blue}.

If the global block sizes are imbalanced, we use Eq.~(\ref{eq:influence}) to adapt the influence values for the next round (Line~\ref{line:adapt-influence}).
As the block sizes were communicated in Line~\ref{line:communicate-sizes}, this can be done independently by each process.
The algorithm returns either when the blocks are sufficiently balanced or when a maximum number of balancing iterations is reached.
In our experiments with $\epsilon \in \{0.03, 0.05\}$, balance was always achieved when allowing a sufficient number of balance and movement iterations.

\begin{algorithm}[thb]
 \KwIn{points $P$, number of blocks $k$, maximum imbalance $\epsilon$, deltaThreshold}
 \KwOut{assignments $A$}
 $n \gets\#P$\;
 $\#$proc $\gets$ number of processors\;
 $r$ $\gets$ rank of processor\;
 sfcIndex[$p$] $\gets$ index of $p$ on space-filling curve $\forall p \in P$ \;\label{line:calculate-sfc}
 \textcolor{markedcolor}{sortedPoints $\gets$ sortGlobal($P$, key=sfcIndex)}\;\label{line:sort-sfc}
 \textcolor{markedcolor}{$P_{\text{local}}$ $\gets$ sortedPoints[$r\cdot n/\#\text{proc}, ..., (r+1)\cdot n/\#\text{proc}$]}\;\label{line:redistribute-sfc}
 \textcolor{markedcolor}{$C[i]$ $\gets$ sortedPoints[$i\cdot n/k + n/2k$] for $i \in \{0, ... k-1\}$}\;\label{line:sfc-centers}
 $I[c]$ $\gets$ 1 for $c \in C$\;\label{line:initialize-influence}
 ub[$p$] $\gets$ $\inf$, lb[$p$] = $0$ $ \forall p\in P_{\text{local}}$\;\label{line:initialize-bounds}

 \For{$i \in \{0, ..., \mathrm{maxIter}\}$}{
  \textcolor{markedcolor}{$A$, $I$, ub, lb $\gets$ AssignAndBalance($C$, $P_{\text{local}}$, $I$, $A$, ub, lb, $\epsilon$)}\;\label{line:call-assign-and-balance}
  $C'_{\text{local}}[c] \gets$ mean of $p \in P_{\text{local}}$ with $A[p] = c$\;
  \textcolor{markedcolor}{$C'$ $\gets$ globalWeightedMeanVector($C'_{\text{local}}[c]$)}\;\label{line:new-centers}
  \If{$\max\delta(C,C') < $ deltaThreshold}{
  \Return{$A$}
  }
  $C \gets C'$\;
  adapt bounds ub, lb with Eq.~(\ref{eq:ub-influence}) and~(\ref{eq:lb-influence})\;
 }
 \Return{$A$}
 \caption{BalancedKMeans}
 \label{algo:balanced-k-means}
\end{algorithm}
\renewcommand{\floatpagefraction}{.6}

\begin{figure*}[tb]
	\includegraphics[scale=0.096]{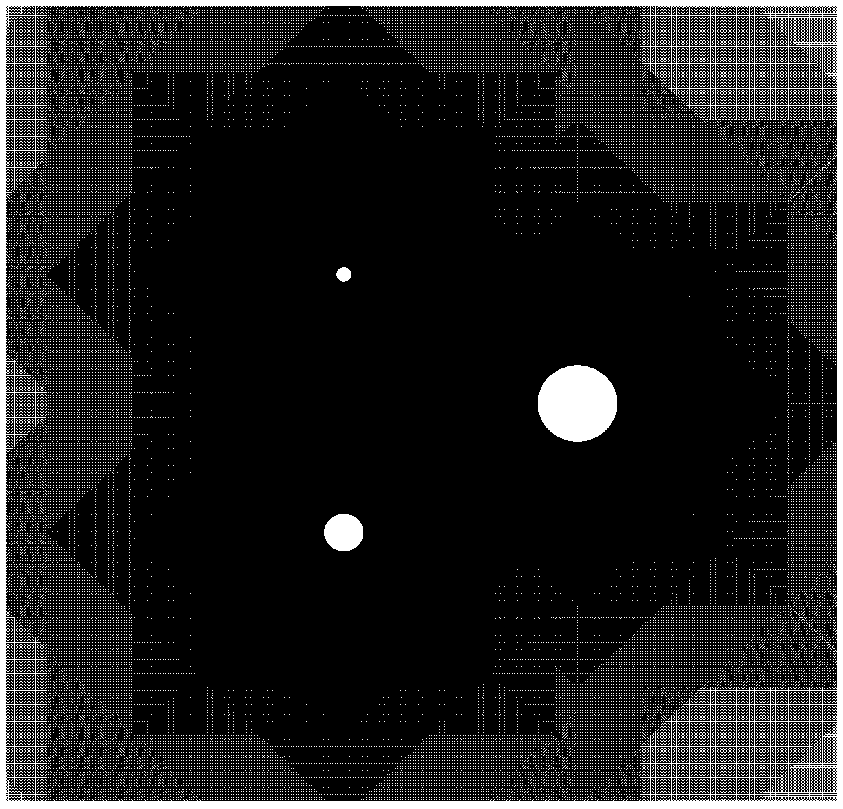}
	\includegraphics[scale=0.1]{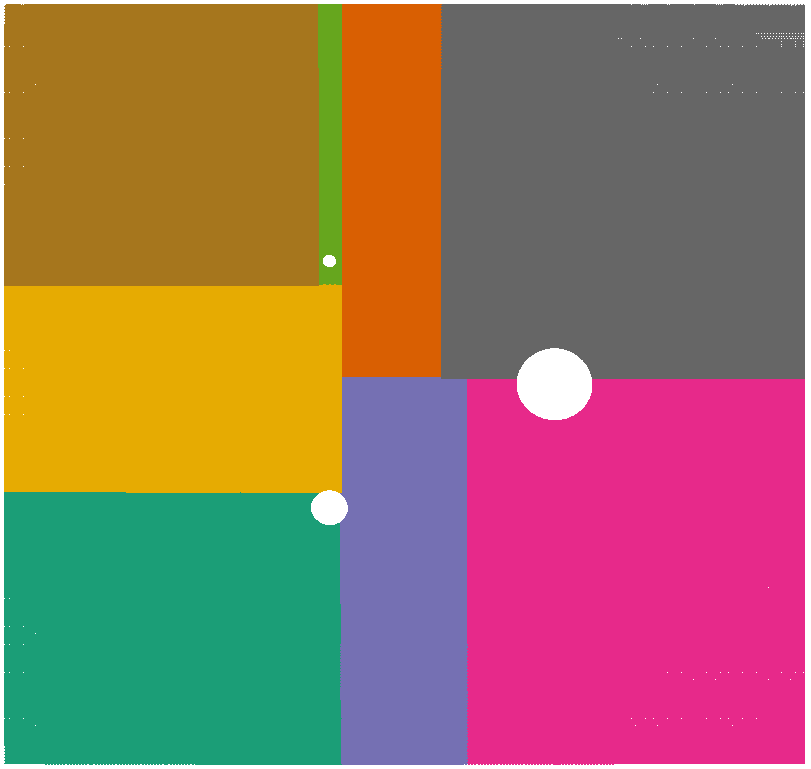}
	\includegraphics[scale=0.1]{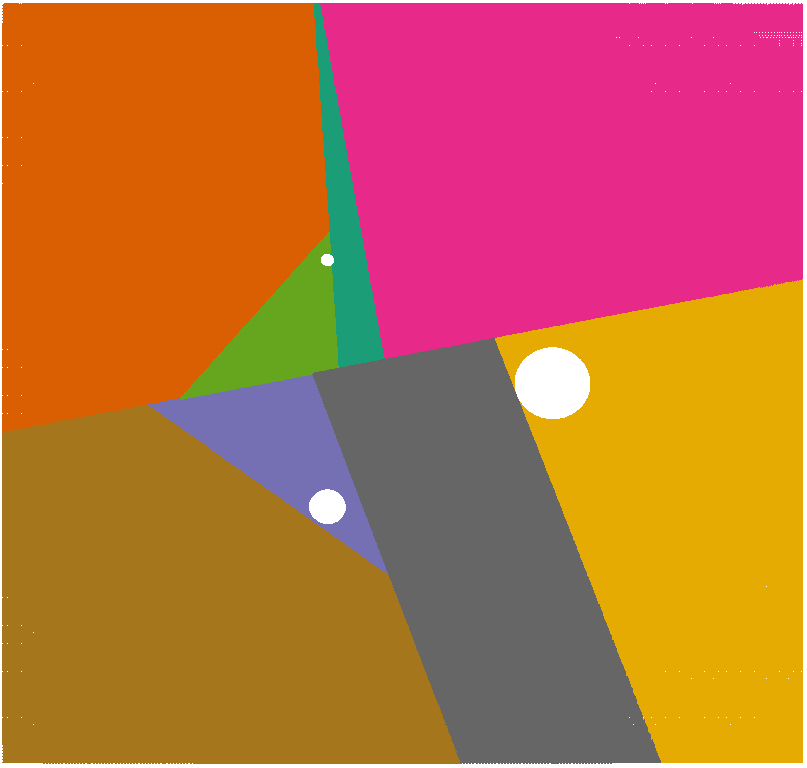}
	\includegraphics[scale=0.1]{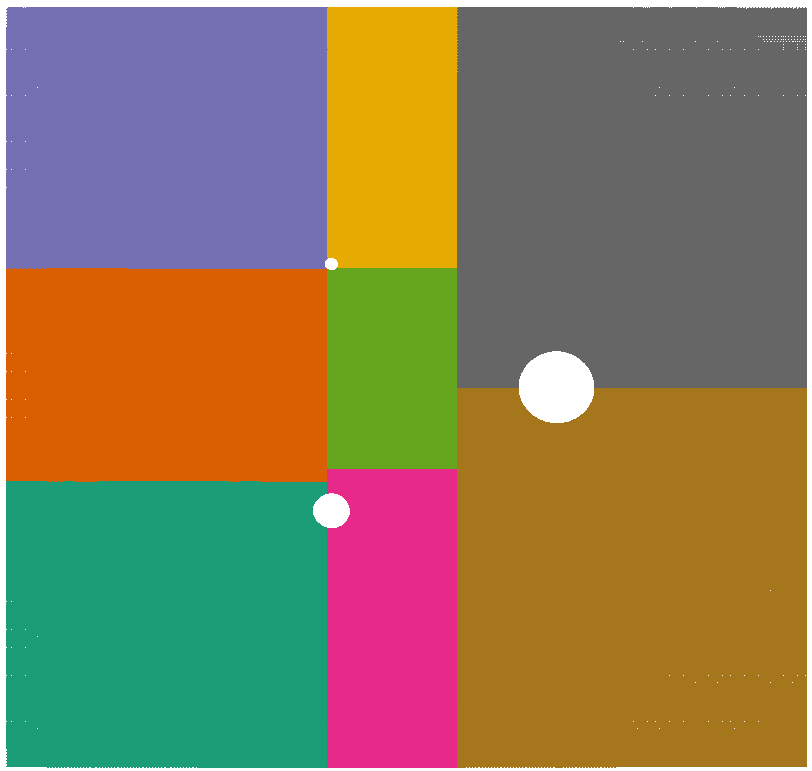}
	\includegraphics[scale=0.1]{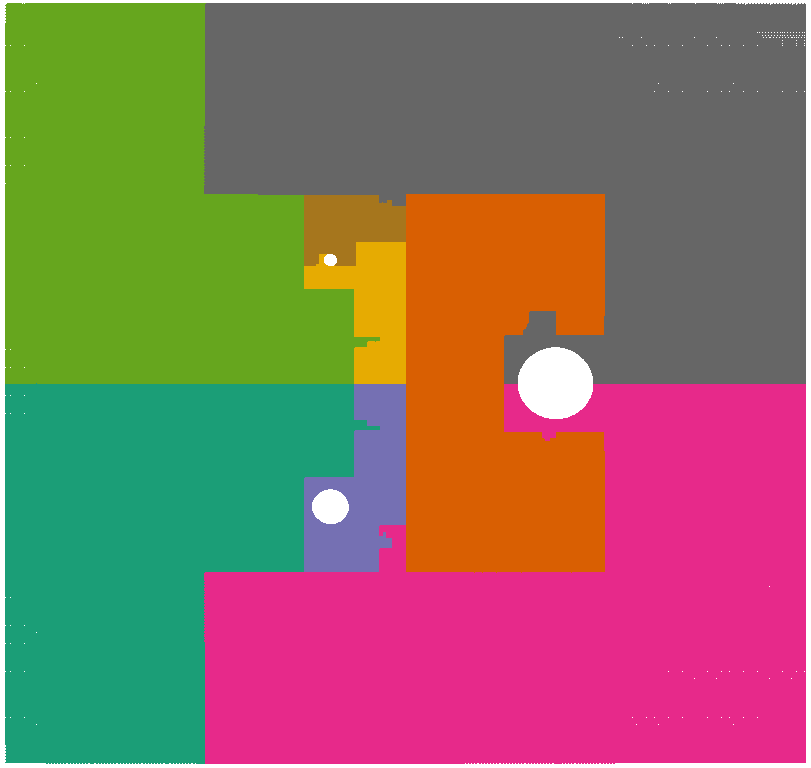}
	\includegraphics[scale=0.1]{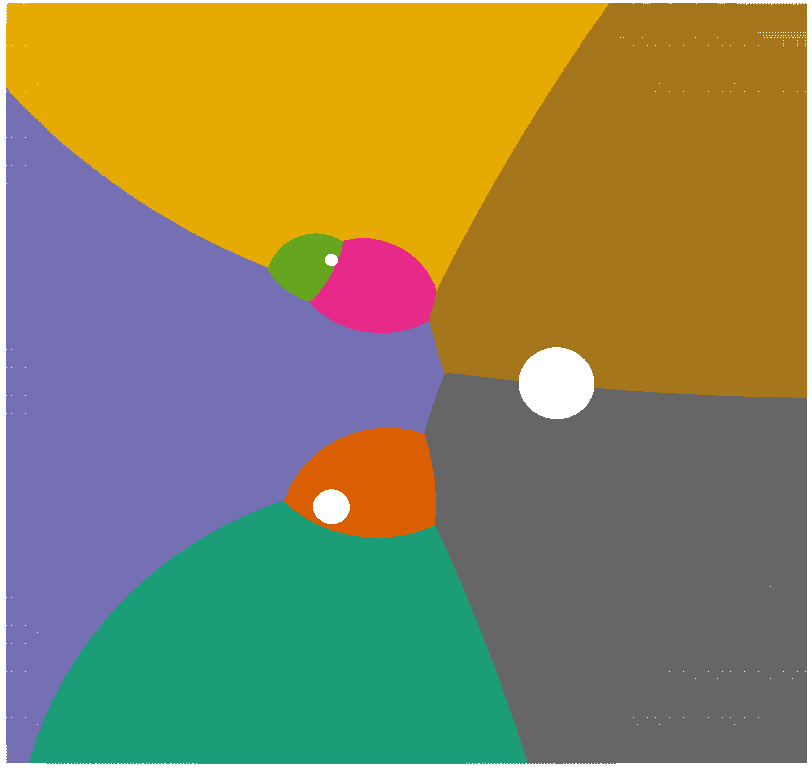}
	\caption{Partition of hugetric-0000 in 8 blocks with different tools. From left to right, the pictures show the input and the results of \rcb, \rib, \mj, \zsfc and \bakpa. }
	\label{fig:hugetric-all-tools}
\end{figure*}

Algorithm~\ref{algo:balanced-k-means} shows our main $k$-means algorithm.
We first sort and redistribute the points according to a space-filling curve to improve spatial locality (Lines~\ref{line:calculate-sfc} - \ref{line:redistribute-sfc}),
place initial centers in equal distances on the sorted points (Line~\ref{line:sfc-centers}) and initialize data structures (Lines~\ref{line:initialize-influence} and \ref{line:initialize-bounds}).
Deriving initial centers from the space-filling curve in this way yields a beneficial geometric spread.

The main loop consists of calling Algorithm~\ref{algo:assign-and-balance} until the centers converge sufficiently or a maximum number of iterations is reached.
New cluster centers are the weighted average of the assigned points; this can be computed efficiently with a local sum and two global MPI vector sum operations (Line~\ref{line:new-centers}).
Apart from the initial setup, all communication steps are global reduction operations, for which efficient implementations exist.
Lines needing communication are marked in \textcolor{markedcolor}{blue}.
Note that the number of blocks the point set is partitioned into is completely independent from the number of parallel processes that are used to do it.

One optimization omitted from the pseudocode for the sake of brevity is \emph{random initialization}.
In the initial phases of $k$-means, cluster centers and influence values change rapidly and the geometric bounds are of little help.
However, during these wild fluctuations not as much precision is required as in the later fine-tuning stages.
To exploit this effect, each process permutes its local points randomly and then picks the first 100 as initial sample.
After each round with center movement, the sample size is doubled.
These $\lceil \log_2 (n_{\text{local}}/100)\rceil$ initialization rounds take about as much time as one round with the full point set, but proceed much further.
Starting with only a randomly sampled subset of points does not impact the quality noticeably.

\section{Experimental Evaluation}
\label{sec:experiments}

\begin{figure*}
\centering
\begin{subfigure}[t]{0.32\textwidth}
\centering
	\begin{tikzpicture}
	\begin{axis}[ylabel= relative value,
	ybar,
	ymin = 0.85,
	yticklabel style = {font=\tiny, xshift=0.5ex},	
	ytick distance = 0.1,
	bar width=2.5pt,
	legend style={at={(1.2,0.7)}},
	xtick=\empty,
	width=1.1\textwidth,
	extra x tick labels={edgeCut, maxCommVol, totCommVol, harmDiam, timeComm},
	extra x ticks={0, 1, 2, 3, 4},
	extra x tick style={grid=major,tick label style={rotate=90} }]
	\addplot [geoKmeans, sharp plot,update limits=false, line width=1pt] plot coordinates{
	(-1,1)
	(5,1)
	};
	\addlegendentry{geoKmeans}
	\addplot [zoltanMJ, fill=zoltanMJ] plot coordinates{
	(0 , 1.30627109716)
	(1 , 1.54590490949)
	(2 , 1.3098160004)
	(3 , 1.36903100283)
 	(4 , 1.02317249203)
	};
	\addlegendentry{zoltanMJ}
	\addplot[zoltanRcb, fill=zoltanRcb] plot coordinates{
	(0 , 1.17827778607)
	(1 , 1.26890478393)
	(2 , 1.17829003583)
	(3 , 1.20424428271)
 	(4 , 1.47725880492)
	};
	\addlegendentry{zoltanRcb}
	\addplot  [zoltanRib, fill=zoltanRib] plot coordinates{
	(0 , 1.1529053978)
	(1 , 1.22321609107)
	(2 , 1.15127818967)
	(3 , 1.26177260985)
 	(4 , 1.49292651796)
	};
	\addlegendentry{zoltanRib}
	\addplot [zoltanHsfc, fill=zoltanHsfc]  plot coordinates{
	(0 , 1.4628259825)
	(1 , 1.58699253863)
	(2 , 1.5339119206)
	(3 , 1.84581799302)
 	(4 , 1.69506505088)
	};
	\addlegendentry{zoltanHsfc}
	\legend{};
	\end{axis}
	\end{tikzpicture}
	\caption{DIMACS graphs (2D)}
\end{subfigure}
\hfill
\begin{subfigure}[t]{0.32\textwidth}
\centering
	\begin{tikzpicture}
	\begin{axis}[
	ybar,
	yticklabel style = {font=\tiny, xshift=0.5ex},
	bar width=2.5pt,
	legend style={at={(0.8,1.32)}},
	ytick distance = 0.1,	
	xtick=\empty,
	width=1.1\textwidth,
	extra x tick labels={edgeCut, maxCommVol, totCommVol, harmDiam, timeComm},
	extra x ticks={0, 1, 2, 3, 4},
	extra x tick style={grid=major,tick label style={rotate=90} }]
	\addplot [geoKmeans, sharp plot,update limits=false, line width=1pt] plot coordinates{
	(-1,1)
	(5,1)
	};
	\addlegendentry{geoKmeans}
	\addplot[zoltanMJ, fill=zoltanMJ] plot coordinates{
	(0 , 0.994373729503)
	(1 , 0.953505815959)
	(2 , 1.06891996464)
	(3 , 0.896783465513)
 	(4 , 1.03798660272)
	};
	\addlegendentry{MJ}
	\addplot[zoltanRcb, fill=zoltanRcb] plot coordinates{
	(0 , 1.04044186839)
	(1 , 1.05960769233)
	(2 , 1.07675100075)
	(3 , 0.917896744401)
 	(4 , 1.4176786646)
	};
	\addlegendentry{Rcb}
	\addplot [zoltanRib, fill=zoltanRib] plot coordinates{
	(0 , 1.11184820433)
	(1 , 1.13902352282)
	(2 , 1.13206720797)
	(3 , 1.1753503648)
 	(4 , 1.46988686152)
	};
	\addlegendentry{Rib}
	\addplot [zoltanHsfc, fill=zoltanHsfc]  plot coordinates{
	(0 , 1.27688839511)
	(1 , 1.17171369644)
	(2 , 1.33955142359)
	(3 , 1.14720595699)
 	(4 , 1.50203979439)
	};
	\addlegendentry{Hsfc}
	\end{axis}
	\end{tikzpicture}
	\caption{Climate graphs (2.5D)}
\end{subfigure}	
%
\begin{subfigure}[t]{0.32\textwidth}
\centering
	\begin{tikzpicture}
	\begin{axis}[
	ybar,
	yticklabel style = {font=\tiny, xshift=0.5ex},
	ytick distance = 0.1,
	bar width=2.5pt,
	xtick=\empty,
	width=1.1\textwidth,
	extra x tick labels={edgeCut, maxCommVol, totCommVol, harmDiam, timeComm},
	extra x ticks={0, 1, 2, 3, 4},
	extra x tick style={grid=major,tick label style={rotate=90} }]
	\addplot [geoKmeans, sharp plot,update limits=false, line width=1pt] plot coordinates{
	(-1,1)
	(5,1)
	};
	\addlegendentry{\bakpa}
	\addplot[zoltanMJ, fill=zoltanMJ] plot coordinates{
	(0 , 0.952312318807)
	(1 , 0.980427405513)
	(2 , 1.0315580851)
	(3 , 1.00794926968)
 	(4 , 1.05121316958)
	};
	\addlegendentry{MJ}
	\addplot [zoltanRcb, fill=zoltanRcb] plot coordinates{
	(0 , 0.988043700645)
	(1 , 0.986662060556)
	(2 , 1.06270094852)
	(3 , 1.09467219324)
 	(4 , 1.25209140031)
	};
	\addlegendentry{Rcb}
	\addplot[zoltanRib, fill=zoltanRib] plot coordinates{
	(0 , 1.06735050061)
	(1 , 1.11034650338)
	(2 , 1.1043073136)
	(3 , 1.21705596062)
 	(4 , 1.3973104355)
	};
	\addlegendentry{Rib}
	\addplot[zoltanHsfc, fill=zoltanHsfc] plot coordinates{
	(0 , 1.10876795628)
	(1 , 1.14506930254)
	(2 , 1.18247674161)
	(3 , 1.18592018576)
 	(4 , 1.49343549878)
	};
	\addlegendentry{Hsfc}
	\legend{};
	\end{axis}
	\end{tikzpicture}
	\caption{Alya and Delaunay (3D) }
\end{subfigure}	
\caption{Aggregated ratios (geometric mean except diameter, see text) of the evaluation metrics for all tools. Baseline: \bakpa.}
\label{plot:quality-comparison}
\end{figure*}
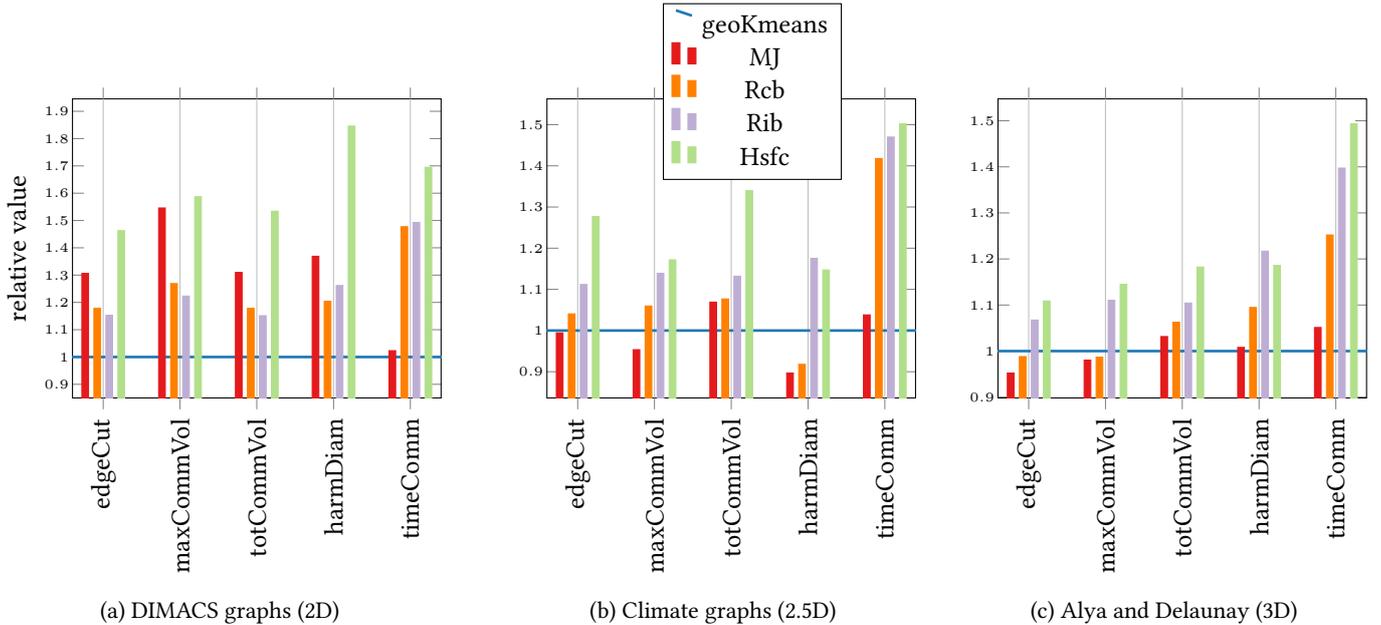

\subsection{Implementation}
Our graph partitioner \bakpa is implemented in C++11 and parallelized with MPI\@.
To increase portability and usability, we develop the partitioner within LAMA~\cite{brandes2017lama}, a portable framework for distributed linear algebra and other numerical applications.
LAMA provides high-level data structures and communication routines for distributed memory, abstracting away the specifics of the MPI communicator and also supporting other parallelization mechanisms.
In the course of this work, we contributed several optimized communication routines to LAMA that are used by our partitioner.

\subsection{Experimental Settings}
\label{sub:exp-settings}
\subsubsection{Machine}
We perform our experiments on Thin Phase 1 nodes of the SuperMUC petascale system at LRZ.
Each node is equipped with 32 GB RAM and two Intel Xeon E5-2680 processors (Sandy Bridge) at 2.7 GHz and 8 cores per processor.
In our experiments, we allocate one MPI process to each core.
Both our code and the evaluated competitors are compiled with GCC 5.4 and parallelized with IBM MPI 1.4.
The repository of our implementation will be made public after paper acceptance.

\subsubsection{Other Partitioners}
We compare \bakpa with several established geometric partitioning implementations from the \parmetis (4.0.3) and \zoltan 2 (part of the Trilinos 12.10 Project) toolboxes. 
The geometric partitioner within the \parmetis package uses space-filling curves, the \zoltan package contains implementations of Recursive Coordinate Bisection (\rcb), Recursive Inertial Bisection (\rib), space-filling curves (\zsfc) and the \mj algorithm, mentioned in Section~\ref{sec:rel-work}.
Since the \parmetis version of space-filling curves is dominated by the space-filling curves in the \zoltan package, we omit it from the detailed presentation.

\subsubsection{Test Data}
We evaluate the partitioners on a variety of datasets:
a collection of benchmark meshes from the 10th DIMACS implementation challenge~\cite{BaderMSW12b-dimacs}, 
2.5D meshes with node weights from the climate simulations~\cite{gmd-10-765-2017} described in Section~\ref{sec:introduction},
3D meshes from the PRACE Unified European Applications Benchmark Suite (UEABS)~\cite{UEABS} and
Delaunay triangulations\footnote{We thank Christian Schulz, who kindly provided the generator~\cite{HoltgreweSS10engineering} for the \graphname{Delaunay} graphs.} of random points in 2 and 3 dimensions.

More precisely, the graphs \graphname{hugetrace}, \graphname{hugetric} and \graphname{hugebubbles} are 2D adaptively refined triangular meshes from the benchmark generator created by Marquardt and Schamberger~\cite{MarquardtSchamberger05open}; they represent synthetic numerical simulations and have approx. 5M to 20M vertices.
\graphname{333SP},  \graphname{AS365}, \graphname{M6}, \graphname{NACA0015} and \graphname{NLR}  are 2D finite element triangular meshes from approx.\ 3.5M vertices and 11M edges up to approx.\ 21M vertices and 31M edges. 
\graphname{rgg\_n} are 2D random geometric graph with $2^n$ vertices for  $n=20,\dots,24$.
 All these graphs come from the 10th DIMACS Implementation Challenge~\cite{BaderMSW12b-dimacs}.
The graphs in the \graphname{DelaunayX} series are Delaunay triangulations of \graphname{X} random 2D points in the unit square~\cite{HoltgreweSS10engineering}.
The smallest graph in the series has 8M vertices and approximately 24M edges; the largest one has 2B vertices and approximately 6B edges.
We also generated five 3D \graphname{Delaunay} triangulations from approx.\ 1M to 16M vertices using the generator of Funke \etal~\cite{funke2018communication}.
The graphs \graphname{alyaTestCaseA} with 9.9 million vertices and \graphname{alyaTestCaseB} with 30.9 million vertices (representing the respiratory system) are from the PRACE benchmark suite~\cite{UEABS}.

\subsubsection{Metrics}
We evaluate the generated partitions with respect to the edge cut, the communication volume, the diameter and the SpMV communication time -- \ie the metrics introduced in Section~\ref{sec:prob-def}.
We report both $\max \mathrm{comm}$, the maximum communication volume and $\sum \mathrm{comm}$, the total communication volume.

To evaluate the effect of shape optimization, we also measure the diameter for each block.
Since a precise computation of the diameter scales at least quadratically with the number of nodes,
we instead use a lower bound generated by executing the first 3 rounds of the iFUB algorithm by Crescenzi et al.~\cite{crescenzi2013computing}.
This lower bound is a 2-approximation of the exact diameter, but often already tight.

\subsubsection{Other Parameters}
In all experiments, we set the number of blocks $k$ to the number of processes $p$ and the maximum imbalance $\epsilon$ to $3\%$, which was respected by all tools.
Reported values are averaged over \numprint{5} runs to account for random fluctuations.

\subsection{Results}  
\label{sub:exp-results}

For a brief first visual impression of the results of \bakpa, \rcb, \rib, \mj, and \zsfc, see Fig.~\ref{fig:hugetric-all-tools}.
Recursive coordinate and inertial bisection produce thin, long blocks, MultiJagged produces rectangles with a better aspect ratio, \zsfc's blocks have wrinkled boundaries, balanced k-means produces curved blocks.

\subsubsection{Quality}
Figure~\ref{plot:quality-comparison} compares the partitions yielded by the tested tools under the metrics edgeCut, maximum and total communication volume and diameter.
For easier presentation, we report the relative value compared to \bakpa and aggregate the results by graph class using the geometric mean.

In some cases, blocks are disconnected and thus have an infinite diameter.
To avoid a potentially infinite mean diameter, we use the harmonic instead of the geometric mean to aggregate the diameter over all blocks.

The first instance class consists of the 2D geometric benchmark meshes from the DIMACS challenge, the second consists of the 2.5D graphs from climate simulations. The third class consists of the alya test case and 
Delaunay triangulations in the unit cube, all 3D meshes.

In all graph classes, \bakpa produces on average the partition with the lowest total communication volume.
The advantage is most pronounced on the 2D geometric meshes from the DIMACS collection, but visible also in other classes.

The performance as measured by the edge cut differs:
On the DIMACS graphs, \bakpa is leading with 15\% difference, on the 2.5D and 3D graphs \tool{MultiJagged} has an advantage of 0.5\% and 4\%, respectively.
Similar developments are visible also for other metrics.
Of course, this does not mean that \bakpa achieves \emph{always} the best results, as these are aggregated values.

Strangely, the empirical average communication time within the SpMV benchmarks (timeComm in Figure~\ref{plot:quality-comparison}) correlates little with the more established measures.
Results fluctuate, but \bakpa has on average the smallest SpMV communication time.

Note that the behaviour of balanced $k$-means is \emph{stable}: If it performs worse on some class, then not by much.
None of the evaluated competitors clearly dominates:
While MultiJagged, for example, has a lower mean edge (5\%) cut on 3D graphs, its performance on the DIMACS graphs is clearly worse, with a 30\% higher edge cut.

Detailed results for individual graphs can be found in Tables~\ref{table:results-large-graphs} and \ref{table:results-small-graphs}.

\begin{small}
\begin{table*}
\caption{Comparison of results for large graphs for $k=p=1024$. Best values are marked in bold.}
   \begin{center}
	\begin{tabular}{ l l l l l l l l l}
	graph name & tool & time & cut & maxCommVol & $\Sigma$ commVol & diameter & timeSpMVComm\\
	\hline
	\graphname{alyaTestCaseB} &\bakpa & \numprint{0.3588} & \numprint{5823055} & \numprint{6508} & \textbf{\numprint{5403716}} & \numprint{63} & \numprint{0.00019877} & \\
	n = \numprint{30959144} &\tool{HSFC} & \numprint{0.0352704} & \numprint{6613710} & \numprint{8275} & \numprint{6802889} & \numprint{83} & \numprint{0.00036143} & \\
	&\mj & \textbf{\numprint{0.0234486}} & \textbf{\numprint{5364660}} & \textbf{\numprint{6062}} & \numprint{5482086} & \textbf{\numprint{62}} & \textbf{\numprint{0.00019804}} & \\
	&\rcb & \numprint{0.0966411} & \numprint{6188060} & \numprint{6526} & \numprint{5825470} & \numprint{79} & \numprint{0.00031597} & \\
	\hline
	\graphname{delaunay250M} &\bakpa & \numprint{0.8308} & \textbf{\numprint{2037960}} & \textbf{\numprint{2183}} & \textbf{\numprint{2033939}} & \textbf{\numprint{457}} & \textbf{\numprint{3.461e-05}} & \\
	n = \numprint{250000000} &\tool{HSFC} & \textbf{\numprint{0.227387}} & \numprint{2356510} & \numprint{2918} & \numprint{2349673} & \numprint{570} & \numprint{0.00013638} & \\
	&\mj & \numprint{0.228118} & \numprint{2118810} & \numprint{2214} & \numprint{2114657} & \numprint{494} & \numprint{7.432e-05} & \\
	&\rcb & \numprint{0.857324} & \numprint{2118220} & \numprint{2211} & \numprint{2113916} & \numprint{491} & \numprint{0.00012676} & \\
	\hline
	\graphname{delaunay2B} &\bakpa & \numprint{6.0989} & \textbf{\numprint{5771443}} & \textbf{\numprint{6136}} & \textbf{\numprint{5754364}} & - & - & \\
	n = \numprint{2000000000} &\tool{HSFC} & \numprint{1.87662} & \numprint{6335780} & \numprint{6807} & \numprint{6314790} & - & - & \\
	&\mj & \textbf{\numprint{1.71619}} & \numprint{5994110} & \numprint{6164} & \numprint{5976573} & - & - & \\
	&\rcb & \numprint{6.66174} & \numprint{5995910} & \numprint{6137} & \numprint{5978340} & - & - & \\
	\hline
	\graphname{fesom-jigsaw} &\bakpa & \numprint{0.7744} & \numprint{33135424} & \numprint{1539} & \numprint{680638} & \numprint{392} & \textbf{\numprint{3.769e-05}} & \\
	n = \numprint{14349744} &\tool{HSFC} & \textbf{\numprint{0.0193754}} & \numprint{33749900} & \numprint{1500} & \numprint{735964} & \numprint{339} & \numprint{5.812e-05} & \\
	&\mj & \numprint{0.019648} & \textbf{\numprint{27472100}} & \textbf{\numprint{1111}} & \textbf{\numprint{641574}} & \numprint{320} & \numprint{4.353e-05} & \\
	&\rcb & \numprint{0.0821395} & \numprint{30447900} & \numprint{1524} & \numprint{642714} & \textbf{\numprint{320}} & \numprint{6.778e-05} & \\
	\hline
	\graphname{refinedtrace-00006} &\bakpa & \numprint{1.5063} & \textbf{\numprint{813450}} & \textbf{\numprint{1596}} & \textbf{\numprint{1380977}} & \textbf{\numprint{1052}} & \textbf{\numprint{4.095e-05}} & \\
	n = \numprint{289383634} &\tool{HSFC} & \numprint{0.51603} & \numprint{1044170} & \numprint{2856} & \numprint{1909341} & \numprint{1834} & \numprint{0.00011955} & \\
	&\mj & \textbf{\numprint{0.260334}} & \numprint{1063020} & \numprint{3790} & \numprint{1801389} & \numprint{1607} & \numprint{7.164e-05} & \\
	&\rcb & \numprint{1.10246} & \numprint{930479} & \numprint{3864} & \numprint{1552315} & \numprint{1335} & \numprint{0.00010835} & \\
	\hline
	\graphname{refinedtrace-00007} &\bakpa & \numprint{3.5417} & \textbf{\numprint{1144423}} & \textbf{\numprint{2205}} & \textbf{\numprint{1948602}} & \textbf{\numprint{1483}} & \textbf{\numprint{5.198e-05}} & \\
	n = \numprint{578551252} &\tool{HSFC} & \numprint{0.812107} & \numprint{1467320} & \numprint{4054} & \numprint{2689631} & \numprint{2609} & \numprint{0.00015608} & \\
	&\mj & \textbf{\numprint{0.570606}} & \numprint{1521740} & \numprint{4644} & \numprint{2551801} & \numprint{2285} & \numprint{8.913e-05} & \\
	&\rcb & \numprint{2.03696} & \numprint{1314980} & \numprint{6988} & \numprint{2189002} & \numprint{1898} & \numprint{0.00014698} & \\
	\end{tabular}
	\end{center}
  \label{table:results-large-graphs}
\end{table*}
\end{small}

\begin{small}
\begin{table*}
   \caption{Comparison of results for small and medium graphs for $k=p=64$. Best values are marked in bold.}
   \begin{center}
	\begin{tabular}{ l l l l l l l l l}
	graph name & tool & time & cut & maxCommVol & $\Sigma$ commVol & diameter & timeSpMVComm\\
	\hline
	\graphname{333SP} &\bakpa & \numprint{1.4991} & \textbf{\numprint{32170}} & \textbf{\numprint{1203}} & \textbf{\numprint{32306}} & \textbf{\numprint{205}} & \numprint{8.166e-05} & \\
	n = \numprint{3712815} &\tool{HSFC} & \textbf{\numprint{0.0545837}} & \numprint{83162} & \numprint{2341} & \numprint{83077} & \numprint{845} & \numprint{6.17e-05} & \\
	&\mj & \numprint{0.0781859} & \numprint{90650} & \numprint{3899} & \numprint{90793} & \numprint{517} & \textbf{\numprint{3.396e-05}} & \\
	&\rcb & \numprint{0.0819943} & \numprint{59558} & \numprint{2291} & \numprint{59700} & \numprint{357} & \numprint{4.576e-05} & \\
	\hline
	\graphname{AS365} &\bakpa & \numprint{0.4749} & \textbf{\numprint{51666}} & \numprint{1114} & \textbf{\numprint{51832}} & \textbf{\numprint{274}} & \numprint{2.965e-05} & \\
	n = \numprint{3799275} &\tool{HSFC} & \numprint{0.0567997} & \numprint{87274} & \numprint{1863} & \numprint{87355} & \numprint{487} & \numprint{6.469e-05} & \\
	&\mj & \textbf{\numprint{0.049213}} & \numprint{64312} & \numprint{1796} & \numprint{64480} & \numprint{364} & \textbf{\numprint{2.174e-05}} & \\
	&\rcb & \numprint{0.101785} & \numprint{55880} & \textbf{\numprint{1075}} & \numprint{56040} & \numprint{313} & \numprint{5.214e-05} & \\
	\hline
	\graphname{M6} &\bakpa & \numprint{0.2921} & \textbf{\numprint{51971}} & \numprint{1145} & \textbf{\numprint{52131}} & \textbf{\numprint{259}} & \numprint{2.771e-05} & \\
	n = \numprint{3501776} &\tool{HSFC} & \numprint{0.0449845} & \numprint{82905} & \numprint{1749} & \numprint{82938} & \numprint{416} & \numprint{6.417e-05} & \\
	&\mj & \textbf{\numprint{0.041304}} & \numprint{58270} & \numprint{1080} & \numprint{58430} & \numprint{310} & \textbf{\numprint{2.289e-05}} & \\
	&\rcb & \numprint{0.0856968} & \numprint{56867} & \textbf{\numprint{1052}} & \numprint{57027} & \numprint{301} & \numprint{5.229e-05} & \\
	\hline
	\graphname{NACA0015} &\bakpa & \numprint{0.1066} & \textbf{\numprint{27841}} & \textbf{\numprint{549}} & \textbf{\numprint{27997}} & \textbf{\numprint{152}} & \numprint{2.426e-05} & \\
	n = \numprint{1039183} &\tool{HSFC} & \textbf{\numprint{0.0125552}} & \numprint{56902} & \numprint{1367} & \numprint{56897} & \numprint{370} & \numprint{5.145e-05} & \\
	&\mj & \numprint{0.0140939} & \numprint{40314} & \numprint{759} & \numprint{40476} & \numprint{244} & \textbf{\numprint{1.941e-05}} & \\
	&\rcb & \numprint{0.031087} & \numprint{29484} & \numprint{603} & \numprint{29638} & \numprint{162} & \numprint{3.16e-05} & \\
	\hline
	\graphname{NLR} &\bakpa & \numprint{0.3744} & \textbf{\numprint{56805}} & \textbf{\numprint{1073}} & \textbf{\numprint{56969}} & \textbf{\numprint{281}} & \numprint{2.893e-05} & \\
	n = \numprint{4163763} &\tool{HSFC} & \textbf{\numprint{0.0548839}} & \numprint{85740} & \numprint{1704} & \numprint{85803} & \numprint{377} & \numprint{6.383e-05} & \\
	&\mj & \numprint{0.0594569} & \numprint{61034} & \numprint{1102} & \numprint{61193} & \numprint{306} & \textbf{\numprint{2.368e-05}} & \\
	&\rcb & \numprint{0.113787} & \numprint{60703} & \numprint{1170} & \numprint{60863} & \numprint{306} & \numprint{4.799e-05} & \\
	\hline
	\graphname{alyaTestCaseA} &\bakpa & \numprint{0.966} & \numprint{894845} & \numprint{18341} & \numprint{841593} & \numprint{113} & \textbf{\numprint{0.00021497}} & \\
	n = \numprint{9938375} &\tool{HSFC} & \numprint{0.122195} & \numprint{988168} & \numprint{21325} & \numprint{1001416} & \numprint{132} & \numprint{0.00035605} & \\
	&\mj & \textbf{\numprint{0.0801415}} & \textbf{\numprint{839540}} & \numprint{17798} & \numprint{839437} & \textbf{\numprint{107}} & \numprint{0.0002506} & \\
	&\rcb & \numprint{0.172015} & \numprint{847188} & \textbf{\numprint{17726}} & \textbf{\numprint{839377}} & \numprint{108} & \numprint{0.00027482} & \\
	\hline
	\graphname{alyaTestCaseB} &\bakpa & \numprint{1.8657} & \numprint{1869558} & \numprint{38203} & \textbf{\numprint{1774168}} & \numprint{163} & \textbf{\numprint{0.00032937}} & \\
	n = \numprint{30959144} &\tool{HSFC} & \numprint{0.395728} & \numprint{1857670} & \numprint{39351} & \numprint{1880792} & \numprint{171} & \numprint{0.00058969} & \\
	&\mj & \textbf{\numprint{0.255547}} & \textbf{\numprint{1772580}} & \textbf{\numprint{37435}} & \numprint{1785528} & \textbf{\numprint{156}} & \numprint{0.00051183} & \\
	&\rcb & \numprint{0.577578} & \numprint{1784790} & \numprint{37447} & \numprint{1785598} & \numprint{167} & \numprint{0.00057401} & \\
	\hline
	\graphname{delaunay017M} &\bakpa & \numprint{0.7332} & \textbf{\numprint{122875}} & \textbf{\numprint{2235}} & \textbf{\numprint{122634}} & \textbf{\numprint{476}} & \textbf{\numprint{3.013e-05}} & \\
	n = \numprint{17000000} &\tool{HSFC} & \numprint{0.249101} & \numprint{131407} & \numprint{2428} & \numprint{131012} & \numprint{549} & \numprint{0.00010383} & \\
	&\mj & \textbf{\numprint{0.17899}} & \numprint{125088} & \numprint{2302} & \numprint{124823} & \numprint{489} & \numprint{3.832e-05} & \\
	&\rcb & \numprint{0.385619} & \numprint{124789} & \numprint{2263} & \numprint{124545} & \numprint{499} & \numprint{0.00010632} & \\
	\hline
	\graphname{fesom-f2glo04} &\bakpa & \numprint{0.4482} & \textbf{\numprint{4758930}} & \textbf{\numprint{1590}} & \textbf{\numprint{66472}} & \textbf{\numprint{547}} & \textbf{\numprint{2.948e-05}} & \\
	n = \numprint{5945730} &\tool{HSFC} & \textbf{\numprint{0.0684101}} & \numprint{7677820} & \numprint{2372} & \numprint{108910} & \numprint{957} & \numprint{4.574e-05} & \\
	&\mj & \numprint{0.0796806} & \numprint{5866330} & \numprint{2263} & \numprint{85789} & \numprint{740} & \numprint{3.185e-05} & \\
	&\rcb & \numprint{0.137691} & \numprint{5596840} & \numprint{1755} & \numprint{79220} & \numprint{631} & \numprint{4.222e-05} & \\
	\hline
	\graphname{fesom-fron} &\bakpa & \numprint{0.5098} & \textbf{\numprint{3870505}} & \numprint{1565} & \textbf{\numprint{61576}} & \numprint{731} & \numprint{3.32e-05} & \\
	n = \numprint{5007727} &\tool{HSFC} & \textbf{\numprint{0.0546617}} & \numprint{5272540} & \numprint{1993} & \numprint{90305} & \numprint{1067} & \numprint{3.205e-05} & \\
	&\mj & \numprint{0.0609416} & \numprint{4214460} & \textbf{\numprint{1485}} & \numprint{70794} & \numprint{788} & \textbf{\numprint{2.691e-05}} & \\
	&\rcb & \numprint{0.0903249} & \numprint{4365920} & \numprint{1846} & \numprint{74330} & \textbf{\numprint{706}} & \numprint{3.441e-05} & \\
	\hline
	\graphname{fesom-jigsaw} &\bakpa & \numprint{1.0854} & \numprint{8444620} & \numprint{4225} & \numprint{166006} & \numprint{6306} & \numprint{4.233e-05} & \\
	n = \numprint{14349744} &\tool{HSFC} & \numprint{0.183937} & \numprint{8224760} & \numprint{4269} & \numprint{177071} & \numprint{2159} & \numprint{0.00012137} & \\
	&\mj & \textbf{\numprint{0.176129}} & \textbf{\numprint{6185490}} & \textbf{\numprint{3078}} & \textbf{\numprint{142214}} & \textbf{\numprint{1843}} & \textbf{\numprint{4.101e-05}} & \\
	&\rcb & \numprint{0.393419} & \numprint{8637620} & \numprint{3980} & \numprint{173583} & \numprint{2800} & \numprint{0.00012333} & \\
	\hline
	\graphname{hugebubbles-00020} &\bakpa & \numprint{2.5848} & \textbf{\numprint{47763}} & \textbf{\numprint{1636}} & \textbf{\numprint{81556}} & \textbf{\numprint{1048}} & \textbf{\numprint{3.1e-05}} & \\
	n = \numprint{21198119} &\tool{HSFC} & \textbf{\numprint{0.28146}} & \numprint{63053} & \numprint{2561} & \numprint{118785} & \numprint{2002} & \numprint{7.507e-05} & \\
	&\mj & \numprint{0.283137} & \numprint{60958} & \numprint{2430} & \numprint{105714} & \numprint{1453} & \numprint{3.191e-05} & \\
	&\rcb & \numprint{0.612614} & \numprint{57482} & \numprint{2003} & \numprint{98404} & \numprint{1299} & \numprint{7.898e-05} & \\
	\hline
	\graphname{hugetrace-00020} &\bakpa & \numprint{1.4833} & \textbf{\numprint{43522}} & \textbf{\numprint{1471}} & \textbf{\numprint{74122}} & \textbf{\numprint{948}} & \numprint{3.102e-05} & \\
	n = \numprint{16002413} &\tool{HSFC} & \numprint{0.20332} & \numprint{50800} & \numprint{2081} & \numprint{98367} & \numprint{1585} & \numprint{5.583e-05} & \\
	&\mj & \textbf{\numprint{0.192364}} & \numprint{50493} & \numprint{1872} & \numprint{86699} & \numprint{1117} & \textbf{\numprint{2.76e-05}} & \\
	&\rcb & \numprint{0.393817} & \numprint{50851} & \numprint{1911} & \numprint{84526} & \numprint{1084} & \numprint{3.904e-05} & \\
	\hline
	\graphname{hugetric-00020} &\bakpa & \numprint{0.8206} & \textbf{\numprint{29298}} & \textbf{\numprint{972}} & \textbf{\numprint{49899}} & \textbf{\numprint{615}} & \textbf{\numprint{2.878e-05}} & \\
	n = \numprint{7122792} &\tool{HSFC} & \numprint{0.157777} & \numprint{39373} & \numprint{1658} & \numprint{72388} & \numprint{1210} & \numprint{5.725e-05} & \\
	&\mj & \textbf{\numprint{0.0928525}} & \numprint{39382} & \numprint{2203} & \numprint{67949} & \numprint{899} & \numprint{3.033e-05} & \\
	&\rcb & \numprint{0.176951} & \numprint{35512} & \numprint{1364} & \numprint{60706} & \numprint{810} & \numprint{5.482e-05} & \\
	\hline
	\graphname{rdg-3d} &\bakpa & \numprint{0.3243} & \textbf{\numprint{1481602}} & \numprint{12683} & \textbf{\numprint{761610}} & \textbf{\numprint{45}} & \textbf{\numprint{0.00025151}} & \\
	n = \numprint{4194304} &\tool{HSFC} & \textbf{\numprint{0.0461232}} & \numprint{1596600} & \numprint{13391} & \numprint{821701} & \numprint{50} & \numprint{0.00032872} & \\
	&\mj & \numprint{0.0573633} & \numprint{1537820} & \textbf{\numprint{12552}} & \numprint{790899} & \numprint{48} & \numprint{0.00027877} & \\
	&\rcb & \numprint{0.0882342} & \numprint{1537980} & \numprint{12558} & \numprint{791086} & \numprint{48} & \numprint{0.00029771} & \\
	\end{tabular}
    \end{center}
    \label{table:results-small-graphs}
\end{table*}
\end{small}

\subsubsection{Scaling and Running Time}

\begin{figure}[tb]
\centering
\begin{subfigure}{\linewidth}
\centering
\begin{tikzpicture}
	\begin{axis}[
	ylabel= seconds, 
	legend style={at={(0.5,1)}},
	xmode = log, 
	ymode=log,
	log basis x= 2,
	xtick={32, 64, 128, 256, 512, 1024, 2048, 4096, 8192},
	line width= 1.5pt]
\addplot plot [geoKmeans, mark options={geoKmeans}] coordinates{	
	(32, 1.1777)
	(64, 0.7332)
	(128, 0.9879)
	(256, 0.9169)
	(512, 0.8486)
	(1024, 0.8308)
	(2048, 0.9861)
	(4096, 1.1534)
	(8192, 1.857)
	};
	\addlegendentry{\bakpa}
	\addplot plot [zoltanMJ, mark options={zoltanMJ}] coordinates{
	(32, 0.167088)
	(64, 0.17899)
	(128, 0.225048)
	(256, 0.225066)
	(512, 0.236598)
	(1024, 0.228118)
	(2048, 0.258429)
	(4096, 0.317737)
	(8192, 0.387967)
	};
	\addlegendentry{MJ}
	\addplot plot [zoltanRcb, mark options={zoltanRcb}] coordinates{
	(32, 0.233472)
	(64, 0.385619)
	(128, 0.498269)
	(256, 0.574832)
	(512, 0.628763)
	(1024, 0.857324)
	(2048, 1.29032)
	(4096, 3.0004)
	(8192, 8.25635)
	};
	\addlegendentry{Rcb}
	\addplot plot [zoltanRib, mark options={zoltanRib}] coordinates{
	(32, 0.289242)
	(64, 0.418331)
	(128, 0.527584)
	(256, 0.648111)
	(512, 0.68487)
	(1024, 0.931585)
	(2048, 1.29113)
	(4096, 2.74664)
	(8192, 7.62787)
	};
	\addlegendentry{Rib}
	\addplot plot [zoltanHsfc, mark options={zoltanHsfc}] coordinates{
	(32, 0.189501)
	(64, 0.249101)
	(128, 0.218983)
	(256, 0.230684)
	(512, 0.222965)
	(1024, 0.227387)
	(2048, 0.285534)
	(4096, 0.392267)
	(8192, 0.493067)
	};
	\addlegendentry{Hsfc}
	\end{axis}
	\end{tikzpicture}
  \caption{Weak scaling for \graphname{DelaunayX} graph series.}
  \label{plot:weak-scal-supermuc-delaunay:time} 
\end{subfigure}
%
%
\begin{subfigure}[t]{0.45\textwidth}
\centering
	\begin{tikzpicture}
	\begin{axis}[
	xlabel=k, 
	ylabel= seconds, 
	legend style={at={(1.5,0.7)}},
	xmode = log, 
	ymode=log,
	log basis x= 2,
	xtick={1024, 2048, 4096, 8192, 16384},
	line width= 1.5pt]
	\addplot plot [geoKmeans, mark options={geoKmeans}] coordinates{	
	(1024, 6.0989)
	(2048, 3.3637)
	(4096, 1.887)
	(8192, 1.857)
	(16384, 2.3413)
	};
	\addlegendentry{\bakpa}
	\addplot plot [zoltanMJ, mark options={zoltanMJ}] coordinates{
	(1024, 1.71619)
	(2048, 0.966196)
	(4096, 0.560776)
	(8192, 0.387967)
	(16384, 0.445028)
	};
	\addlegendentry{zoltanMJ}
	\addplot plot [zoltanRcb, mark options={zoltanRcb}] coordinates{
	(1024, 6.66174)
	(2048, 4.35643)
	(4096, 4.00006)
	(8192, 8.25635)
	(16384, 24.6478)
	};
	\addlegendentry{zoltanRcb}
	\addplot plot [zoltanRib, mark options={zoltanRib}] coordinates{
	(1024, 7.44088)
	(2048, 4.72432)
	(4096, 4.11693)
	(8192, 7.62787)
	(16384, 23.2226)
	};
	\addlegendentry{zoltanRib}
	\addplot plot [zoltanHsfc, mark options={zoltanHsfc}] coordinates{
	(1024, 1.87662)
	(2048, 0.921476)
	(4096, 0.60041)
	(8192, 0.493067)
	(16384, 0.64755)
	};
	\addlegendentry{zoltanHsfc}
	\legend{};		
	\end{axis}
	\end{tikzpicture}
	\caption{Strong scaling for \graphname{Delaunay2B} graph}
	\label{plot:strong-scaling-supermuc-delaunay:time} 
\end{subfigure}	

\caption{Scaling results on \graphname{DelaunayX} regarding (a) strong and (b) weak scaling.}
\label{plot:scaling-supermuc-delaunay:time}
\end{figure}
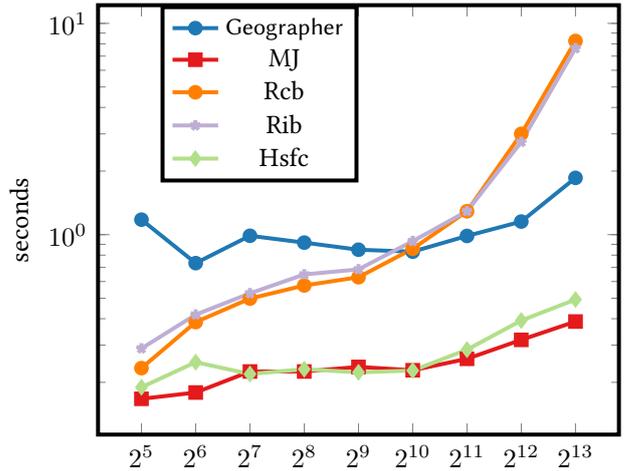
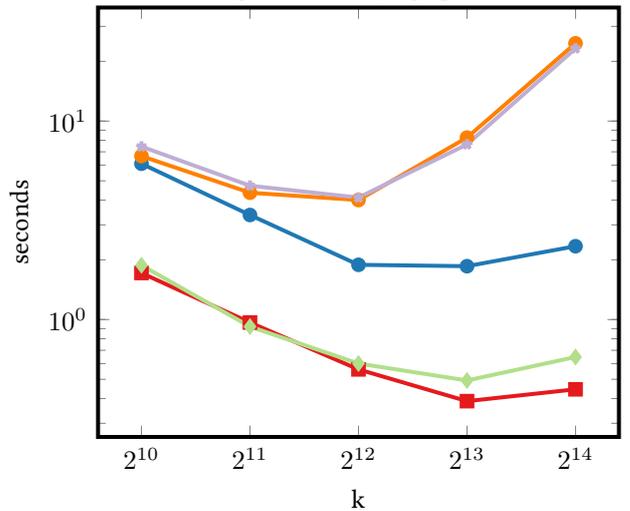

\begin{figure}[tb]
\centering
\begin{tikzpicture}
 \begin{axis}[xmode=log,ymode=log, log basis x = 2, log basis y = 2, xlabel=n,ylabel=seconds,legend entries={}, legend style={at={(0.8,1.70)}}]
  \addplot[scatter,only marks,
	   point meta = explicit symbolic,
	   scatter/classes={
	    geoKmeans={draw=geoKmeans,fill=geoKmeans },%
	    zoltanRcb={draw=zoltanRcb,fill=zoltanRcb },%
	    zoltanRib={draw=zoltanRib,fill=zoltanRib},
	    zoltanMJ={draw=zoltanMJ,fill=zoltanMJ},
	    zoltanHsfc={draw=zoltanHsfc, fill=zoltanHsfc}}
	    ]
	    table[meta=tool]
          {time-scatter.dat};
  \addlegendentry{\bakpa};
  \addlegendentry{\rcb};
  \addlegendentry{\rib};
  \addlegendentry{\mj};
  \addlegendentry{\tool{HSFC}};
  \addplot[dashed,geoKmeans] expression[domain=2^22:2^31] { x^(-0.00894563)*2^(0.99935072)};\addlegendentry{\bakpa, least squares fit};
  \addplot[dashed,zoltanRcb] expression[domain=2^22:2^31] { x^(0.37450822)*2^(-10.05832285)};\addlegendentry{\rcb, least squares fit};
  \addplot[dashed,zoltanRib] expression[domain=2^22:2^31] { x^(0.36163548)*2^(-9.62074319)};\addlegendentry{\rib, least squares fit};
  \addplot[dashed,zoltanMJ] expression[domain=2^22:2^31] { x^(0.10452987)*2^(-4.55235804)};\addlegendentry{\mj, least squares fit};
  \addplot[dashed,zoltanHsfc] expression[domain=2^22:2^31] { x^(0.17426604)*2^(-5.9443695)};\addlegendentry{\tool{HSFC}, least squares fit};
 \end{axis}
\end{tikzpicture}
\caption{Comparison of running times. Each dot represents the running time of one tool on one graph.
For comparable results, we aimed for \numprint{250000} points per block, selecting the number of blocks $k$ (and processes $k) $ separately for each graph.
Since some tools only support powers of two for the number of blocks, we select the power of two which results in the number of local points closest to our target of \numprint{250000}.
}
 \label{plot:time-scatter}
\end{figure}
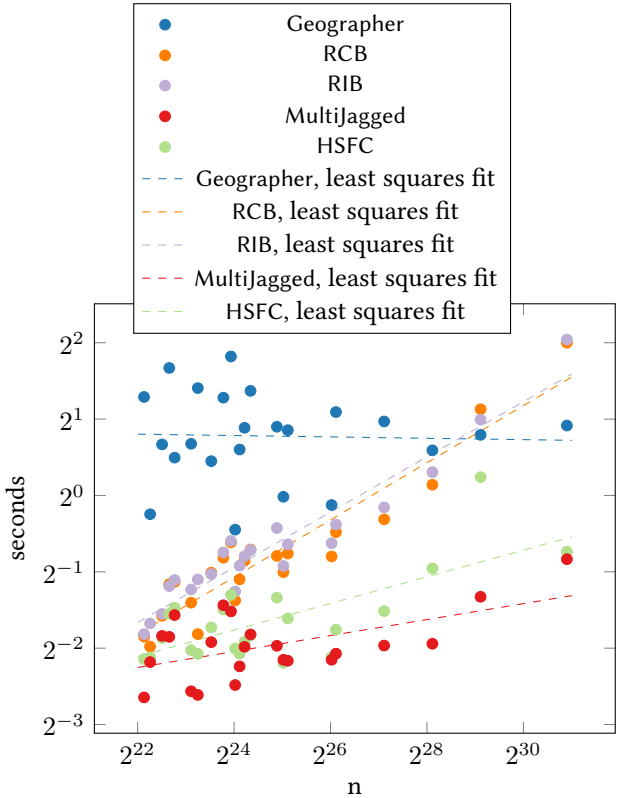

\paragraph{Weak Scaling}
Fig.~\ref{plot:weak-scal-supermuc-delaunay:time} shows a direct comparison of weak scaling performance on the \graphname{DelauanyX} graph series. 
We start with 32 processes ($p$) and blocks ($k$) on 8 million vertices and repeatedly double both until we reach \numprint{8192} processes on 2 billion vertices, keeping the ratio fixed at approx \numprint{250000} vertices per process. 
\bakpa exhibits similar behavior as \mj and \zsfc: they scale almost perfectly up to 1024 PEs, then increase roughly by a factor of two over the next three doublings.
The recursive methods \rib and \rcb show an immediate increase in running time with larger inputs and scale 
especially poorly on more than \numprint{1024} processes, more than doubling the running time for each doubling of input and process count.
A comparison of running times on all graphs can be seen in Fig.~\ref{plot:time-scatter}.
The scaling behavior of the different tools follows similar trends as on the \graphname{DelaunayX} series.
Fitted trend lines show a better weak scaling behavior of \bakpa than on \graphname{Delaunay} graphs alone,
which may also be an artifact of the higher number of smaller graphs in our collection.

\paragraph{Strong Scaling}
We perform strong scaling experiments (see Fig.~\ref{plot:strong-scaling-supermuc-delaunay:time}) with the largest graph at our disposal, \graphname{Delaunay2B}.
With 2 billion vertices, it is small enough to fit into the memory of \numprint{1024} processing elements but also sufficiently large to partition it into \numprint{16384} blocks.
Note that our experiments are not, strictly speaking, strong scaling, as we increase the number of blocks along with the number of processors.

Similarly to the weak scaling results, \bakpa, \mj and \zsfc have similar scaling behavior: almost perfect scalability for up to \numprint{4096} PEs (\mj also for \numprint{8192}).
\rcb and \rib start with the slowest running times, around $6.5$ seconds for $k=1024$ and climb to $23$ seconds for $k=16384$ showing poor scalability.
For all tools, the running time  increases from \numprint{8192} to \numprint{16384} processes;
we attribute this to the SuperMUC architecture: an island in SuperMUC contains \numprint{8192} cores and communication is more expensive across islands.

\paragraph{Components}
The main parts of \bakpa contributing to the running time are the initial partition with a Hilbert curve,
the redistribution of coordinates according to this initial partition and finally the balanced $k$-means itself.
As the number of processes increases, the relative share of these components changes:
For small instances, the computation of Hilbert indices and the balanced $k$-means iterations constitute a majority of the time,
while for higher number of processes, the redistribution step dominates.
For example, when partitioning \graphname{Delaunay2B} with 1024 processes, data redistribution and $k$-means take $32\%$ respective $47\%$ of the time.
For the same graph and 16384 processes, redistribution takes $46\%$ and $k$-means $42\%$ of the total running time.

\section{Conclusions}
\label{sec:conclusions}

We designed and implemented a balanced, scalable version of  $k$-means for partitioning geometric meshes.
Combined with space-filling curves for initialization, it scales to thousands of processors and billions of vertices, partitioning them in a matter of seconds.

An evaluation on a wide range of input meshes shows that the total communication volume and resulting SpMV communication time of the resulting partitions is on average 5-15\% better than those of state-of-the-art competitors.
This difference is most pronounced on meshes from the DIMACS benchmark collection, but also measurable on graphs from climate simulations and 3D meshes.
Concerning the edge cut, another common metric to evaluate graph partitioners, \mj also performs well, giving the best results on 3D meshes.
No partitioner dominates on all point sets.

Future work will be concerned with further improving quality and scalability, in particular for 3D meshes.
A faster redistribution routine, necessary to achieve scalability for a higher number of processes, is also of independent interest.
Finding high-quality embeddings of non-geometric graphs into some geometric space in a scalable manner is promising, too.
This preprocessing would allow to apply \bakpa to non-geometric graphs as well.

\section*{Acknowledgement}
This work is partially supported by German Research Foundation (DFG) grant ME 3619/2-1 (TEAM)
and by the German Federal Ministry of Education and Research (BMBF) project WAVE (grant 01$\vert$H15004B).
Large parts of this work were carried out while the authors were affiliated with Karls\-ruhe
Institute of Technology. We gratefully acknowledge the Gauss Centre for Supercomputing e.V. (www.gauss-centre.eu) for funding this project by providing computing time on the GCS Supercomputer SuperMUC at Leibniz Supercomputing Centre (LRZ, www.lrz.de).
We thank Michael Axtmann for providing us with a preliminary implementation of his scalable sorting algorithm.
We thank Vadym Aizinger for supplying simulation meshes from climate simulations and helpful discussions.
We further thank Michael Axtmann, Thomas Brandes and Christian Schulz for helpful discussions.


\bibliographystyle{plain}
\bibliography{refs-parco}
\end{document}